# Self-Hybridized Polaritonic Emission from Layered Perovskites

*Surendra B. Anantharaman,[1] Christopher E. Stevens,[2,3] Jason Lynch,[1] Baokun Song,[1] Jin Hou,[4,5] Huiqin Zhang,[1] Kiyoung Jo,[1] Pawan Kumar,[1] Jean-Christophe Blancon,[4] Aditya D. Mohite,[4,5] Joshua R. Hendrickson,[3] Deep Jariwala[1,\*]*

[1] Department of Electrical and Systems Engineering, University of Pennsylvania, Philadelphia, PA 19104, USA.

[2] KBR, Inc., Beavercreek, OH 45431.

[3] Air Force Research Laboratory, Sensors Directorate, Wright-Patterson Air Force Base, Columbus, OH 45433, USA.

[4] Department of Chemical and Biomolecular Engineering, Rice University, Houston, TX 77005, USA.

[5] Department of Materials Science and Nanoengineering, Rice University, Houston, TX 77005, USA.

E-mail: dmj@seas.upenn.edu

**Abstract**

Light-matter coupling in excitonic materials has been the subject of intense investigation due to emergence of new excitonic materials. Two-dimensional layered hybrid organic/inorganic perovskites (2D HOIPs) support strongly bound excitons at room-temperatures with some of the highest oscillator strengths and electric loss tangents among the known excitonic materials. Here, we report strong light-matter coupling in Ruddlesden-Popper phase 2D-HOIPs crystals without the necessity of an external cavity. We report concurrent occurrence of multiple-orders of hybrid light-matter states via both reflectance and luminescence spectroscopy in thick (> 100 nm) crystals and near-unity absorption in thin (< 20 nm) crystals. We observe resonances with quality factors > 250 in hybridized exciton-polaritons and identify a linear correlation between exciton-polariton mode splitting and extinction coefficient of the various 2D-HOIPs. Our work opens the door to studying polariton dynamics in self-hybridized and open cavity systems with broad applications in optoelectronics and photochemistry.

**Introduction**

Low-dimensional or quantum-confined semiconductors exhibit strong excitonic resonances at room temperatures owing to the weak electric field screening between photo-excited electrons and holes. As a consequence, their optical density of states is focused into narrow spectral resonances. These resonances, when coupled with optical cavity resonances, results in the formation of hybrid states of light and matter termed exciton-polaritons. Polariton states are conventionally formed by coupling semiconductor materials with external optical cavities, either dielectric or plasmonic, that confine photons. Controlling and engineering this optical dispersion has become a frontier topic of research for taming light propagation at the nanoscale. Such possibility has opened new avenues in basic photonic sciences such as generation of artificial lattices of interacting excited quantum states (1) as well as applications such as optical switches (2,3), neuromorphic computing (4), and new photochemistry approaches (5). However, the mandate of using an external cavity to hybridize excitons with photons often limits direct access to them for electrical control or sensing, limiting their further study and applicability.

Coupling light into excitonic semiconductors has been the subject of intense study over the past two decades. Until the advent of 2D semiconductors most excitonic semiconductors were $sp^2$ carbon based or organic in nature (e.g. organic small molecules, polymers, and carbon nanotubes) (6). Carbon, being a low Z element, limits the refractive index of organic semiconductors. Therefore, studies involving strong light-exciton coupling in these semiconductors required confining them into external optical cavities such as plasmonic antennas (7) and dielectric Bragg mirrors (8,9). With the advent of 2D Mo and W chalcogenides and Pb halides that exhibit strong excitonic resonances concurrently with high refractive index, there is renewed opportunity in studying light-exciton coupling in these systems without the need for external optical cavities.  2D hybrid organic-inorganic perovskites (2D HOIPs) are multilayered, naturally occurring semiconducting quantum wells (QWs) composed of inorganic lead halide layers (n - number of $PbI_2$ layers) separated by organic ligands have a direct bandgap combined with high exciton binding energy (<0.5 eV) at room temperature (10). Further, they also possess some of the highest known exciton oscillator strengths and loss tangents (comparable to laser dye media) which makes them an interesting subject for strong light-exciton interaction studies as well as investigation of polariton fluids and in photochemistry. However, investigation of self-trapping of light by 2D HOIPs remains limited despite these extraordinary optical constants. Fundamental investigations of self-trapping of photons in an optical medium are critical both from the perspective of non-linear optics as well as in applications such as LEDs, photovoltaics and lasers. While several studies have been performed to investigate strong light-exciton coupling in 2D HOIPs using external cavities such as Bragg mirrors (11,12) the extraordinary optical constants of 2D HOIPs (13) are sufficient to induce such light-exciton coupling merely with control over crystal thickness; thereby opening unique opportunities for thickness tunable light-exciton coupling and overall optical response including density of optical states (DOS) in 2D HOIPs. Such tunability of optical DOS without an encapsulating cavity medium in direct-gap highly luminescent semiconductors is unexplored and presents new opportunities in photochemistry as well as opto-electronics using strongly coupled light-matter states.

In this study, we experimentally demonstrate a single-mode hybrid state formation in exfoliated 2D HOIP flakes without the necessity of an external cavity. We observe concurrent formation of multiple orders of exciton-polariton quasiparticles in both reflectance and luminescence spectroscopy at room temperatures in 100-150 nm thick Ruddlesden-Popper Phase 2D HOIPs. We further observe that both the strength of coupling and order number of exciton-polariton resonances are a function of crystal height suggesting thickness tunable DOS. We correlate our experimental observations with electromagnetic wave simulations and report close agreement between the two. We further demonstrate that our results are general across all 2D HOIPs and demonstrate that the exciton-polariton splitting is linearly dependent on the degree of quantum confinement in 2D HOIPs of varying n. Finally, using temperature, pump intensity and time dependent measurements we determine the mechanism of energy transfer from hybrid exciton-polariton states to triplet states and report occurrence of Boser action in 2D HOIPs (14).

## Results and Discussion

**Demonstrating light-matter interaction in HOIP flakes on Au substrate**

Our concept of hybrid state formation by strong light-matter interactions stems from exploiting the lossy Fabry-Pérot (FP) like cavity modes present in any material with large losses (15,16). In excitonic materials, this occurs at the excitonic resonance resulting is enhanced quality factors (Q) of the FP like modes. The relatively large Q of these modes at high n and k values results in strong-coupling with intrinsic excitonic modes resulting in the formation of exciton-polaritons. Thick HOIP flakes exfoliated on a Au substrate manifest intrinsic, lossy FP-like cavity modes, while thin HOIP flakes show purely excitonic optical response (Figure 1A). Here, thick refers to a thickness of ~150 nm which is 4x the λ/4n (where λ is wavelength and n is the real refractive index) which is the minimum necessary thickness to attain 180° phase shift to induce thin-film interference. As shown schematically for n=1 (hereafter RP1, same applies for RP2 (n=2) etc.), thin HOIP flakes (<20 nm) show only excitonic nature in absorption and photoluminescence (PL). However, thick HOIP flakes (~100-200 nm) on a Au substrate show a single-mode hybrid state that emerges in both absorption and PL. To demonstrate this concept, we characterize the optical properties of thin and thick flakes of HOIP using reflectance and PL spectroscopy. First, mechanically exfoliated HOIP flakes were directly placed on the template-stripped Au substrate using the well-known scotch-tape route (16). A representative optical image of HOIP (RP2) flakes on the Au substrate used in our experiment shows different optical contrast between the thin and thick RP2 flakes under the optical microscope (Figure 1B). Optical contrast between different flakes emerge from the FP resonance, which was helpful to record thickness-dependent far-field reflectance and PL spectra (16). The thicknesses of thin HOIP flakes determined using atomic force microscopy (AFM) studies for RP1, RP2, and RP3 are ~10 nm, 15 nm, and 10 nm, respectively (Figure 1C, 1D, and Figure S1, Supporting Information). With increasing order of HOIPs, the attenuance recorded from these thin HOIP flakes shows only excitonic absorption with a gradual decrease in attenuance, i.e., near-unity absorption (RP1) to less than 50% (RP3) (Figure 1E). The experimental attenuance data (solid lines) shows excellent qualitative and good quantitative overlap with the theoretical attenuance spectra from TMM calculations (dotted lines) suggesting the accuracy of dielectric functions and the calculation method. Therefore, the light absorption efficiency from different HOIPs is directly attributed to the dielectric functions of the material. It is notable to mention that the extraordinary optical constants of 2D HOIPs in these thin flakes of RP1 results in near-unity (~97% at the exciton resonance) light absorption for ~15 nm thick semiconductor samples which is similar to prior observations in TMDCs and significant still since 2D

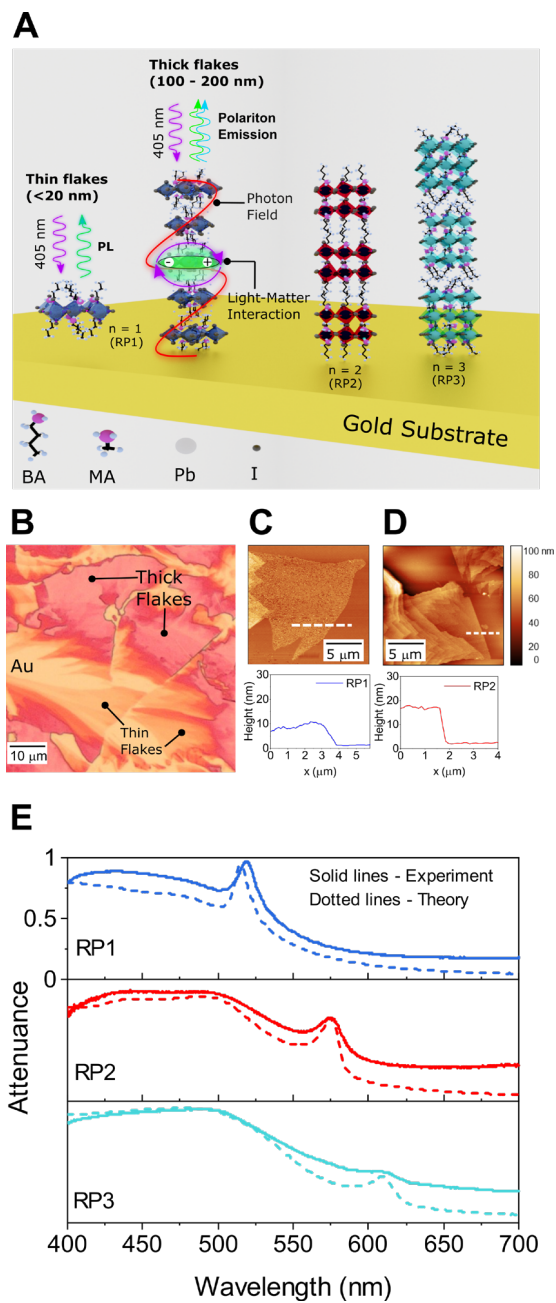

**Fig 1. Exfoliated HOIP flakes on Au substrate demonstrating light-matter interaction.** (**A**) Schematic of RP1 flakes (n=1) on a Au substrate showing photoluminescence from thin RP1 flakes (<20 nm) and hybrid exciton-polariton formation in thick flakes (100-200 nm). A similar phenomenon was also observed in higher-order systems - RP2 (n =2) and RP3 (n = 3). (**B**) Representative optical image of RP2 showing thin and thick flakes on a Au substrate. Atomic force microscopic images from RP1 and RP2 flakes on Au substrates with approximate thicknesses of 10 nm (**C**) and 15 nm (**D**), respectively. The line profile (white dotted line) from the topography is shown below their corresponding images. (**E**) Attenuance (solid line) from RP1 (10 nm) flake showing near-unity absorption at the excitonic wavelength (520 nm) compared to the different thicknesses from higher HOIP system - RP2 (15 nm) and RP3 (10 nm). Dotted lines are attenuance spectra from the transfer-matrix calculations.

HOIPs are direct band gap in nature (17). This light-trapping in ultrathin flakes also suggests great utility in ultrathin photovoltaics and photodetectors made using 2D HOIPs.

**Correlation between Rabi splitting ($\hbar\Omega$) and extinction coefficient ($k$)**

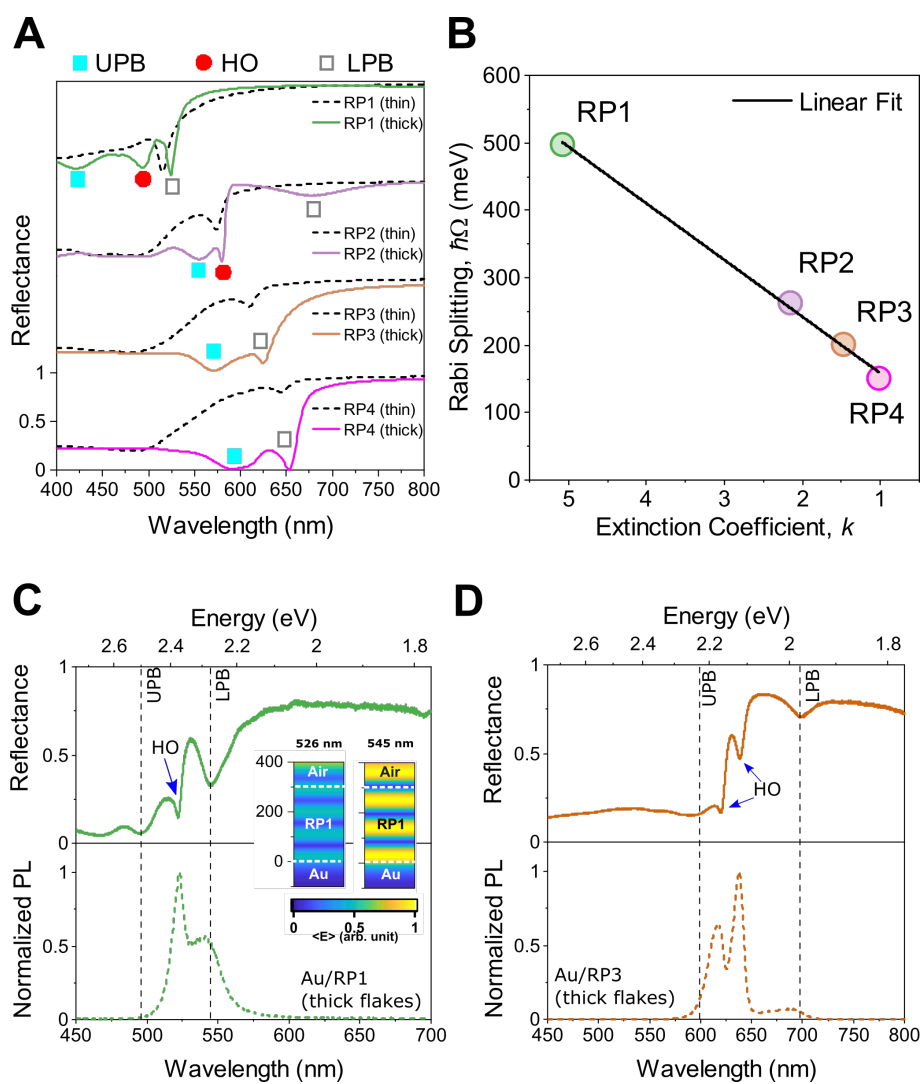

**Figure 2. Correlation between Rabi splitting ($\hbar\Omega$) and extinction coefficient ($k$).** (**A**) Transfer-matrix calculation of reflectance spectra for all HOIP systems (RP1 to RP4) with the optical constant obtained from spectroscopic ellipsometry method. The thin flake thickness was ~10 nm for all RP systems. For hybrid states, maintaining almost similar flake thickness leads to constant mode volume for all the HOIP orders (RP1 ~110 nm and RP2, RP3, and RP4 ~150 nm each) for zero detuning condition. See FigureS2, Supporting Information. Reflectance from thick HOIP flakes shows a clear exciton transparency region in the calculated spectra. (**B**) Theoretical Rabi splitting shows a linear relation with the extinction coefficient of the HOIP ($\hbar\Omega \propto k$). See Supporting Information, Section 1 for details. (**C**) Experimental observation of reflectance data shows strong Rabi splitting in Au/RP1 (300 nm) with higher-order (HO) modes in the reflectance. Inset shows the electric field distribution for the peaks from 526 nm and 545 nm showing HO mode and LPB peak, respectively, from simulation with RP1 thickness ~365 nm. (**D**) Experimental reflectance and PL spectra from Au/RP3 thick flakes (~600 nm) - cavity detuned condition showing the upper polariton branch (UPB), HO mode, and lower polariton branch (LPB) with smallest Rabi splitting. The dotted lines in panel (**C**) and (**D**) corresponds to the reflectance spectra.

Next, we investigate strong coupling in 2D HOIP flakes of varying orders using TMM calculations and correlate our findings with the dielectric functions of the material. First, we will discuss the formation of hybrid states at the zero detuning condition to identify the critical thicknesses of HOIP flakes followed by criteria for HO mode formation. Optical dispersion from different flake thicknesses were analyzed to identify the critical HOIP thickness manifesting in single-mode hybrid state formation. From the simulations, the reflectance spectra for thick flakes (~100-200 nm) show two or more reflectance dips (solid line), in contrast with thin flakes (<20 nm) that only show a single excitonic peak (dotted line) for all HOIPs orders (Figure 2A). We confirmed the presence of a hybrid state (exciton-polariton) in the thick HOIP flakes by fitting the two-coupled oscillator model to the calculated reflectance spectra with increasing HOIP thickness up to 200 nm (Figure S2 a-d, Supporting Information). Varying the flake thickness allows for tuning of the cavity resonance – analogous to varying a distributed Bragg reflector (DBR) microcavity length. Further, from the two-coupled oscillator model, we have calculated the Hopfield coefficients expressed as,

$$\begin{pmatrix} E_x & g \\ g & E_c(t) \end{pmatrix} \begin{pmatrix} \alpha \\ \beta \end{pmatrix} = E(t) \begin{pmatrix} \alpha \\ \beta \end{pmatrix} \quad (1)$$

where $E_x$ is exciton energy, $E_c(t)$ is cavity mode as a function of the thickness (t), and g is the coupling strength. Hopfield coefficients are the value of $|\alpha|^2$ and $|\beta|^2$ that correspond to the exciton and photon component, respectively, from the polariton branch. From our calculation, an equal contribution (or fraction) of exciton and photon was observed at the anticrossing point for all HOIPs (Figure S2e, Supporting Information).

Observing an anticrossing between two modes in the theoretical fitting and a clear exciton transparency region at the excitonic wavelength are unambiguous signatures confirming hybrid exciton-polaritons in thick HOIP flakes. We ascribe these two modes to the upper polariton branch (UPB) and lower polariton branch (LPB). The wavelength for UPB (solid squares in cyan) and LPB (open squares) obtained from the two-coupled oscillator fitting coincides with the reflectance dips in the thick HOIP flakes (RP1 to RP4) (Figure 2A). Our observation of hybrid state formation from thick HOIP flakes in the reflectance spectra is similar to previous reports, which confirmed hybrid states from investigating angle-resolved energy vs. in-plane momentum (11) or oscillation modes in the PL spectra (18) from HOIP flakes without an external cavity. In the absence of an external optical cavity, self-hybridization of the FP cavity modes from the thick flakes in resonance with the excitonic mode facilitates a single-mode hybrid state. Hybrid states formed in flakes with thickness ~100 - 200 nm is critical to maintain similar mode volume for comparing among HOIPs possessing different dielectric functions and eliminate any multi-mode dispersions, previously reported for thick HOIP flakes (600-3000 nm) sandwiched in a dielectric Bragg cavity (19), and on glass substrates (11). Therefore, observing hybrid state formation using lower HOIP thickness is of paramount importance to form condensates in self-hybridized excitonic systems, which is yet to be demonstrated.

In addition to hybrid state formation, we also observe higher-order modes (HO, marked by solid circles in red) present between the UPB and LPB modes. HO modes appear prominently in RP1 and RP2 systems; however, they disappear for RP3 and RP4 due to their low extinction coefficient for zero detuning condition. With increasing flake thickness (non-zero detuned condition), RP3 and RP4 also shows HO modes as discussed later. This contrasting observation is due to the remarkably peaked dielectric functions in RP1 and RP2, which can confine light in the semiconductor medium (discussed later). Nevertheless, hybrid states were observed in all HOIP systems with thicknesses less than 200 nm as evident from the optical simulation.

Rabi splitting is the energy difference between the UPB and LPB at the zero detuning condition ($E_{cavity} = E_{exciton}$), and it was determined for the HOIP systems presented here using the two-coupled oscillator model discussed above. The value of Rabi splitting was plotted against their corresponding extinction coefficient obtained from the ellipsometry data (Fig 2B). A linear fit correlates the Rabi splitting to the absorption coefficient of HOIPs used in this study. The linear correlation can be understood by equating the condition for strong coupling and the Lorentz oscillator model. For strong coupling, the Rabi splitting ($\hbar\Omega$) is a function of coupling strength ($g$), which is related to extinction coefficient ($k$) as,

$$g = k\sqrt{\frac{E_x \gamma_x}{2\epsilon m V_m}} \qquad (2)$$

Here, exciton energy ($E_x$), exciton damping factor ($\gamma_x$), free electron mass ($m$), the real frequency-dependent permittivity of the material ($\epsilon$), and mode volume ($V_m$) correspond to the HOIPs used in this study. From equation (2), it is evident that the Rabi splitting should be linearly proportional to the extinction coefficient of the HOIP, as observed in Figure 2B (refer Section 1, Supporting Information for detailed derivation). It can be noted that RP1 shows the highest Rabi splitting reported so far (11,18,19), while RP4 shows the lowest. Our experimental investigations will focus on hybrid states in RP1 to RP3 as extreme cases and then focus on emission behavior in RP2.

Thick RP1 flakes (300 nm) on Au substrates (Au/RP1) were investigated for hybrid state formation using reflectance and PL spectroscopy at room temperature. The reflectance spectrum for Au/RP1 thick flakes shows hybrid states with UPB and LPB amounting to a Rabi splitting of ~363 meV, almost two-fold (three-fold) higher as compared to previous reports in open cavity (DBR) systems at room temperatures (11,12,20,21). Further, we can experimentally observe HO modes in the reflectance spectra interspersed between the exciton-polariton branches. Visualizing the electric field distribution within the Au/RP1 (300 nm) stack for reflectance wavelengths filtered at 526 nm and 545 nm, as shown in the inset in Figure 2C, shows that the 545 nm traps 1.5 wavelengths of light inside of the RP1 flake which is consistent with LPB modes. Likewise, the reflectance spectrum from a Au/RP3 thick flake (600 nm) shows both hybrid state with $\hbar\Omega$ ~90 meV and HO modes (Figure 2D). Emission from the hybrid state is dominated by the UPB and HO mode as evident from higher PL intensity compared to the LPB. Collectively, the difference between RP1 and RP3 for observing the HO mode corresponds to the real part of the refractive indices of the two crystals, which necessitates higher flake thickness for RP3 compared to RP1.

**Emergence of hybrid state formation and emission from Au/RP2**

To understand the evolution of a single-mode hybrid state at a critical HOIP thickness, we will focus on thickness-dependent reflectance spectroscopy studies in Au/RP2 and then extend our argument for other HOIP systems. Various thicknesses of RP2 flakes exfoliated on a Au substrate were chosen to record the reflectance spectrum at room temperature (Figure 3A). Thin RP2 flakes (~6 nm) show a primary (ground-state) exciton mode at 573.9 nm in the reflectance spectrum. Further, the sharpness of the ground- state exciton peak is due to the transition between the valence and conduction bands only at the $\Gamma$ point. Additional dips in reflectance on the higher energy side (<500 nm) are ascribed to excited exciton states and inter-band transition inherent in HOIP (22,23). With increasing flake thickness (~67 nm), the exciton mode blueshifts then vanishes as an exciton transparency region becomes evident. Peak splitting in the reflectance emerges due to strong coupling between exciton and surface plasmon resonance from the gold surface, as observed in TMDCs before (17). We assign this exciton-plasmon coupling as plexcitons ($\hbar\Omega$ =219 meV), in contrast, to the exciton-polaritons as discussed later. Upon further increasing the flake thickness (110 nm), new peaks emerge at 502.2 nm and 576.5 nm, owing to the UPB (black line) and LPB (red line), respectively. The RP2 flake thickness in the range of 110-150 nm is the zero-detuning condition for exciton-polaritons, as plexciton oscillations can be attenuated at such high dielectric thickness (24). We confirm these mode assignments from the two-coupled oscillator model employed in our simulation studies, as detailed below.

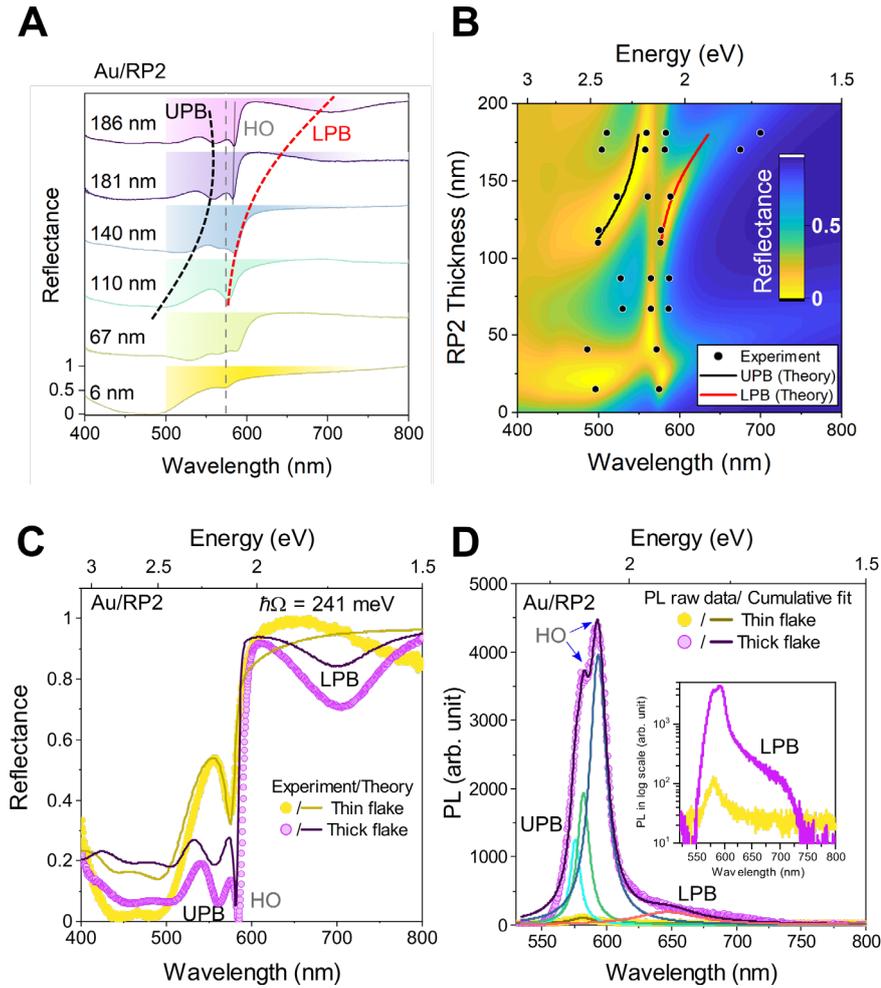

**Figure 3. Emergence of hybrid state formation and emission from Au/RP2.** (**A**) Reflectance spectra from selected thicknesses of RP2 flakes on the Au substrate. With increasing flake thickness, a clear mode splitting (upper polariton branch, UPB and lower polariton branch, LPB) and exciton transparency region (dotted gray line) around 575 nm were observed, confirming the hybrid state formation. The reflectance spectrum from 110 nm thick RP2 flakes shows the UPB and LPB at zero detuning condition. HO modes (solid line) are experimentally observed at higher flake thicknesses. (**B**) 2D false-color map of calculated reflectance spectra for varying RP2 thickness up to 200 nm. The experimental data in (**A**) along with additional spectra recorded from several flake thickness is overlapped as scatter plots. The false color reflectance map from higher RP2 thicknesses are shown in Figure S5, Supporting Information. (**C**) Reflectance spectra for thin (15 nm) and thick (181 nm) RP2 flakes on Au substrate normalized to a silver mirror. Clear anticrossing with energy difference between UPB and LPB of ~241 meV was observed in thick RP2 flakes. The experimental data (scatter) overlaps both qualitatively and quantitatively with the theoretical (line) spectra confirming the hybrid state formation. (**D**) PL peak splitting upon hybrid state formation confirms the emission from the hybrid state. The solid lines are from cumulative fits of Lorentzian function for the exciton mode in thin flakes and UPB, HO modes, and LPB in thick flakes. Inset showing the PL in logarithmic scale where the LPB emission is clearly visible.

TMM calculations were performed to interpret and correlate our experimental observations. Simulation of thickness-dependent reflectance map from Au/RP2 complements and confirms all our experimental data shown in filled circles (Figure 3B). Starting with the RP2 thickness (~15 nm) – exciton and higher exciton states are observed. With an increase in thickness, the exciton mode blueshifts and transitions to an exciton transparency region allowing for the formation of hybrid states in RP2 flakes less than 200 nm thick. Using a two coupled-oscillators model, the fitting obtained from UPB (black line) and LPB (red line) follow closely with our experimental data (filled circles). From the above discussion, we can confirm the formation of a single-mode hybrid state with RP2 thickness less than 200 nm on the Au substrate. In comparison, we investigated similar conditions on a quartz substrate instead of Au. Reflectance and PL spectroscopy from two different flake thicknesses (~30 nm and 232 nm) of RP2 on the quartz substrate were recorded (Supporting Information, Figure S3). Both the experiment and calculations show only exciton broadening, signifying weak light-matter coupling of RP2 flakes on the quartz substrate. These contrasting results in strong coupling on Au substrate and weak coupling on quartz substrate for the same flake thickness emphasize the role of light confinement strategies in sub-wavelength thicknesses. Further, to eliminate any artifact, such as absorption from the Au substrate, we calculated the absorption from individual layers – RP2 and Au (Supporting Information, Figure S4). The absorption from hybrid states was prominently present in the RP2 flakes, and the Au substrate introduced no artifact for thicknesses >50 nm. From the above discussion, we can confirm that the Au substrate plays a vital role in confining light in the HOIP flakes at low thicknesses similar to TMDCs (16,17).

Nevertheless, increasing the HOIP flake thickness beyond 200 nm does not strongly modulate either the Rabi splitting (243 meV to 251 meV) or the UPB peak position, owing to a slight increase in the cavity mode coupling. However, the light-matter interaction manifests into multi-mode LPB from hybridization between the exciton mode and the FP cavity modes in RP2. Similar observations on multi-mode LPB from angle-dependent reflectance and PL was reported upon increasing the HOIP flake thickness in a DBR cavity (19) or on a glass substrate (11). This observation can be applied across all HOIPs from RP1 to RP4 (Supporting Information, Figure S6).

Another mode emerging within the critical thickness of the HOIP flakes (<200 nm) is the higher-order (HO) mode. Comparing thickness-dependent reflectance spectra shows that the HO mode appears close to the exciton branch (Figure 3C). The HO mode originates from the narrowing of the LPB in the higher flake thickness that disperses towards the hybrid states present in the lower thicknesses (modes). Further, the HO mode can be discerned from the hybrid modes by comparing the linewidths. The narrow linewidth of the HO mode, in contrast to the broad linewidth in the hybrid modes, heuristically confirms the signature of exciton-polaritons and the mode assignment. The presence of the exciton-polariton state was also investigated from the PL measurement in thick RP2 flakes. At room temperature, PL emanating from the thick flakes show hybrid emission with dominance in the UPB and HO branch; however, thin flakes show only excitonic emission (Figure 3D). The solid lines are a cumulative fit of four Lorentzian functions used to deconvolute the UPB, HO and LPB emission from the thick flakes and one Lorentzian function for the exciton mode from the thin flakes. It can be noted that the separation of PL peaks between UPB and HO is closer, where UPB emission is blue-shifted from the exciton peak position. The LPB emission is at lower energy ~650 nm, which is clearly visible in the logarithmic scale as shown in the inset Figure 3D. The PL peak splitting between UPB and LPB is comparable to the reflectance spectra, unlike other excitonic materials in plasmonic nanocavities that show very small PL mode splitting (25,26). Based on the above reasons, we do not attribute peak splitting in thick flakes toward electric and magnetic dipole emission, where the emission is polarization-dependent (27) or PL peak splitting owing to exciton fine-structure at low temperatures (6 K) (28).

**Temperature-dependent exciton-polariton behavior from Au/RP2**

We now investigate the temperature-dependent exciton-polariton emission and reflectance spectroscopy from the thick RP2 flakes. We compare our measurements with the exciton behavior from the thin RP2 flake. As we cool down the sample from room temperature (296 K) to 80 K, the PL from thin flakes shows a blueshift around 273 K, due to phase transformation in the butylamine-based halide perovskites used in this study (29,30). A concomitant blueshift in the exciton peak from the reflectance spectra was also observed (Figure 4A and B). At 80 K, the thin flakes show higher absorption and narrower linewidth than at room temperature. PL from hybrid states (HO mode and weakly emitting LPB) in thick flakes also shows a similar blueshift and narrowing of the linewidth upon cooling (Figure 4C). Nevertheless, the single-mode hybrid state emission is retained throughout the thermal excursion to 80 K (Supporting Information, Figure S7).

Hybrid states (UPB and LPB) marked in the reflectance spectra were also vivid from the temperature-dependent reflectance data (Figure 4D). The UPB and LPB follow a similar trend (blueshift) of exciton peak upon cooling. For self-hybridized polaritons, in this case, the cavity mode frequency remains unchanged similar to plasmon resonances during cooling (31). Conversely, we can say that the exciton-polaritons observed at low temperatures are sustained at room temperature. A small peak in the PL emerging at 173 K is confirmed to be from LPB and not to be from biexciton emission as reported previously (12). A corresponding reflectance from LPB is observed at room temperature and even during cooling (Figure 4E). Collectively, we have observed a phase transformation induced blueshift in exciton and hybrid state emission in thick flakes from RP2. Similar exciton-polariton dynamics were also confirmed in RP1 flakes during cooling (Supporting Information, Figure S8).

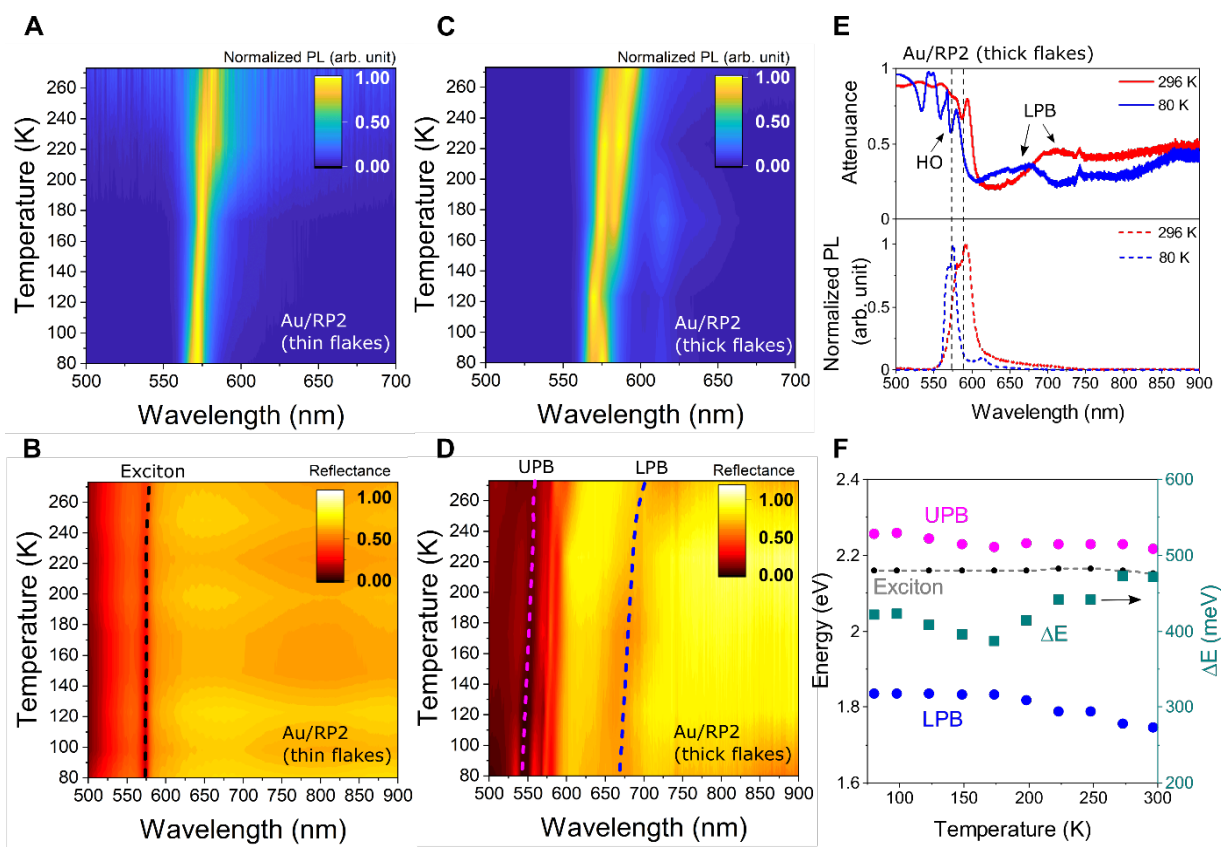

**Figure 4. Temperature-dependent exciton-polariton behavior for Au/RP2.** Normalized PL (**A**) and reflectance (**B**) from thin flakes (25 nm) showing a phase transition around 220 K with a blueshift in peak position. Normalized PL (**C**) and reflectance (**D**) from thick flakes (~180 nm) showing hybrid peaks (HO modes and LPB) undergoing similar blueshift upon cooling to 80 K. PL line spectra are shown in the Supporting Figure S7. The exciton peak position in thin flakes (**B**) and hybrid states (UPB and LPB) from thick flakes (**D**) are shown as dotted lines in the reflectance map. Upon cooling, an enhancement in absorption for thin (**B**) and thick flakes (**D**) can be observed. (**E**) Selected attenuance and PL (296 K and 80 K) from thick flakes showing blueshift and narrowing of PL upon cooling. Conversely, the low temperature PL splitting is maintained at room temperature signifying a room temperature stable hybrid state emission from thick flakes. Peak splitting in higher order (HO) modes are observed at 80 K in the reflectance spectrum. Dotted line shows the exciton attenuance peak from thin flakes. (**F**) Temperature dependent UPB and LPB obtained from reflectance spectra for thick RP2 flakes show a clear transparency region from the exciton peak energy for thin RP2 flakes on the Au substrate. The energy difference (ΔE) between UPB and LPB is shown as filled squares.

Another mode - the HO mode, with prominent peak splitting in the simulations, was obscure from the reflectance spectrum recorded experimentally at room temperature. Upon cooling to 80 K, peak splitting in the HO modes are clearly observed signifying an increase in the oscillator strength (Figure 4E and Supplementary Figure S5). Further, the PL peak splitting in hybrid states straddles the exciton emission wavelength (dotted lines) even at 80 K (see Supplementary Figure S7). For clarity, we have summarized the UPB, LPB, and exciton position from temperature dependence reflectance, which we discussed above (Figure 4F). As the hybrid state formation in the thick flake (180 nm) which has a non-zero detuning energy (close to 125 nm as obtained from the simulations), we have mentioned the energy difference (ΔE) between UPB and LPB, instead of Rabi splitting to avoid overestimation. From these results, we can confirm that exciton-polariton states are present even at low temperatures (80 K) from self-hybridized thick RP2 flakes.

**Exciton-Polariton Dynamics from Time-resolved Measurements at 80 K**

PL recorded from thin RP2 flakes (<20 nm) shows a broad peak at room temperature, as discussed above. Upon cooling to 80 K, exciton fine-structure emerges as observed in previous studies (Supporting Information, Figure S9a) (28). Time-resolved photoluminescence (TRPL) was recorded by selecting different wavelength ranges in the PL signal by introducing band-pass filters (Supporting Information, Figure S9b). Using biexponential fitting, two decay components, fast ($t_1$) and slow ($t_2$), can be obtained by fitting the decay trace. The amplitude of $t_1$ (>95%) was found to be dominating $t_2$ (~5%) in the decay trace fitting. Therefore, the exciton lifetime was in sub-nanosecond (0.27 ns) scale corresponding to the fast component ($t_1$) as reported previously (30). Also, the exciton lifetime from thin flakes was found to be independent of the temperature (Supporting Information, Figure S9d).

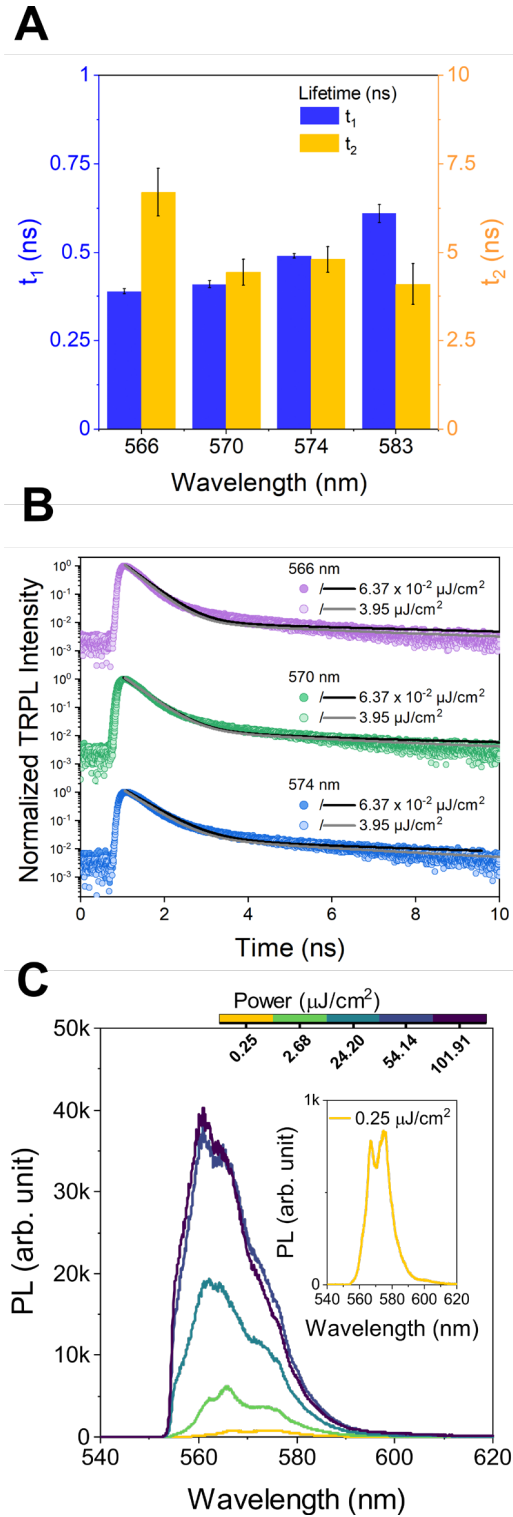

**Figure 5. Time-resolved PL and excitation power dependence studies from hybrid states in RP2 at 80 K.** (**A**) Lifetime from fast ($t_1$) and slow ($t_2$) component at different wavelength such as UPB (565 nm), HO modes - (570 nm and 574 nm) and low energy filtered peak (580 nm). The instrument response function (fast component - 0.015 ns, slow component - 0.27 ns). With increase in PL filtered wavelength, it can be noted that $t_1$ slows down (i.e., increase in lifetime) and $t_2$ becomes faster. (**B**) Corresponding TRPL trace for low ($6.37 \times 10^{-2}$ μJ/cm$^2$) and high (3.95 μJ/cm$^2$) power

shows that lifetime is independent of excitation power used in this study. (**C**) PL spectra with increasing excitation power (mentioned in the color bar) shows a threshold like behavior in the higher energy PL (565 nm), which dominates the lower energy PL (574 nm). Inset shows the PL recorded from low power (0.25 µJ/cm$^2$) excitation.

As the LPB emission is weak for thick flakes compared to UPB and HO mode even at 80 K, we focus our further studies on understanding the emission behavior only from UPB and HO. TRPL measurements were recorded from exciton-polariton emission observed from the thick RP2 flakes at 80 K. Using a biexponential fit, the $t_1$ and $t_2$ obtained from the TRPL trace recorded over different wavelength ranges in the hybrid emission are shown in Figure 5A. The slow component ($t_2$) exhibits an order of magnitude longer lifetime than the fast component ($t_1$). Nevertheless, the $t_1$ from hybrid state emission is similar to the excitonic emission (0.27 ns, Supplementary Material, Figure S9). It can be noted that the $t_1$ ($t_2$) component becomes slower (faster) with an increase in the PL wavelength. We can exclude photon recycling for this observation, as no change in the rise time between the lower and higher wavelength was observed (Figure 5B). Previous works have associated the origin of the long lifetime (differing by an order of magnitude compared to short lifetime) with triplet state emission in HOIPs (32,33). On the other hand, Auger recombination was reported to be a dominant non-radiative decay channel upon increasing the excitation power in RP2 (34). We investigated excitation power dependent TRPL measurements to clarify the mechanism in RP2. The TRPL decay trace showed no change in the lifetime (both $t_1$ and $t_2$) with increasing the excitation power up to 3.82 µJ/cm$^2$ (3000 nW) (Supporting Information, Figure S11). This observation is in perfect agreement with the funneling of excitons from singlet state to triplet state (32) and, therefore, eliminates the role of Auger recombination.

In the case of photon lasing, triplet state was reported for inhibiting lasing (or lasing death) in lower-order HOIPs. Introducing triplet management strategies to suppress the influence of triplet states towards population inversion have led to photon lasing in quasi-2D perovskites (33). Further, continuous wave excitation has shown a two-orders of magnitude lower threshold for photon lasing than pulsed excitation at room temperature. Polariton lasing, on the other hand, follows a two-threshold lasing mediated by biexciton (50 µJ/cm$^2$) followed by condensate formation at a higher threshold regime (600 µJ/cm$^2$) at 4 K (12). In our study, no biexciton emission was observed beyond the polariton emission wavelength in the PL spectrum (Figure 5C). We recorded PL vs. excitation fluence to understand the polariton dynamics in the thick flake at 80 K. At low excitation power (0.25 µJ/cm$^2$), the emission from HO was dominant compared to the UPB emission (inset in Figure5C). With an increase in the excitation fluence, the UPB intensity becomes dominant and blueshifts (558 nm). Collectively, the crossover in dominant peak emission, PL blueshift, and threshold-like PL behavior in UPB emission upon increasing the excitation power are several observations pointing towards a Boser action (14). Boser action sets in when the FP cavity resonance mode is slightly above the exciton resonance mode. In this situation, we can envisage the manifestation of polariton lasing close to the UPB, given that no biexciton lasing was observed. However, more needs to be done to confirm this including excitation at high fluences and maintaining sample photostability at those fluences which is beyond the scope of the present study.

**Conclusion**

In summary, our results highlight a simple and effective approach for light-matter coupling in sub-wavelength thickness (~150 nm) perovskites placed on a gold substrate. This approach led to achieving single-cavity mode polaritons with two-fold higher Rabi splitting than previous reports, which used very thick HOIP flakes (11,12). Our results highlight that achieving strong Rabi splitting in open cavities will be foundational to leverage our understanding of mapping the polariton dynamics using near-field spectroscopy, provided the photostability of the perovskite is enhanced. Strong emission from UPB and HO mode, compared to LPB, may negate biexciton lasing. Further, direct access to semiconductors (HOIP) for electrode contact will benefit electrically driven polariton lasers, polariton photochemistry, and neuromorphic computing.

**Experimental Section**

**Synthesis and growth of 2D HOIP**

2D hybrid organic-inorganic perovskites with Ruddlesden-Popper phase (n = 1 to 3) were synthesized by following the standard route as reported previously (35). Perovskite crystals were grown on a glass substrate from a freshly prepared perovskite solution. 10 µL of the perovskite solution was drop cast on the glass substrate and covered with another glass substrate to allow solvent evaporation overnight at 60 °C. After drying, thick perovskite crystals grown on the glass substrates were used for exfoliation.

**Sample preparation**

Template-stripped Au substrates were prepared by depositing two-component epoxy over the 100 nm Au film thermally evaporated on Si substrate. The epoxy was covered with a Silicon carrier wafer and flipped after drying at 100 °C (16). Using the scotch-tape method, the HOIP flakes (n = 1 to 3) exfoliated from single crystals were placed on the Au substrate inside the glove box.

**Characterizations**

**Steady-state optical measurements**

Samples were placed in a Linkam stage and pumped to $10^{-3}$ Torr. The vacuum level was maintained throughout the measurement. Reflectance spectroscopy was performed in a Horiba LabRam HR Evolution confocal microscope with an external white light source illumination at normal incidence on the sample, and the reflected light was collected through a 50x objective lens (Olympus SLMPLN, N.A. = 0.35). A polished silver mirror was used as a reference to normalize the reflectance from the sample and to avoid any influence of Au absorption in the reflectance spectra of the sample. Photoluminescence (PL) was recorded using a 405 nm continuous-wave excitation under low fluence (6 nW) regime controlled by external neutral density filters to minimize photobleaching. The emitted photons were passed through the 600 grooves/mm grating (Thorlabs) and focused on the spectrometer integrated with a Si focal plane array detector. Temperature-dependent PL and reflectance were recorded using a Linkam stage with a liquid nitrogen supply while cooling and heating, respectively. Thermal excursion between room temperature and 80 K with cooling/heating rate ~5 °C/min and equilibration for 5 min at the desired temperature was maintained before recording the PL and reflectance. The PL spectra were deconvoluted with a Lorentz fit to resolve the peaks.

**Time-resolved PL (TRPL) measurements**

Samples loaded in a Helium cryostat were maintained at 80 K. PL was recorded from thin and thick HOIP flakes at room temperature (296 K), 223 K, and 80 K with 467 nm excitation. The emitted light was transmitted through a 1200 grooves/mm grating to record the PL. Using tunable filters, the PL was filtered at different wavelengths and fiber coupled into a Micro Photon Devices single photon detector. The signal from the detector and the reference from the laser source was fed into a PicoQuant HydraHarp 400 timing box to record the lifetime. For these measurements, a picosecond excitation source at 467 nm with 100 MHz repetition rate was used to excite the sample. The setup provided an instrument response time of 0.015 ns ($t_1$) and 0.2 ns ($t_2$). TRPL was recorded with increasing excitation power from 6.37 x $10^{-2}$ µJ/cm$^2$ to 3.95 µJ/cm$^2$ from thick flakes where the hybrid state is prominently observed.

**Atomic force microscopy**

Topography scans performed using OmegaScope Smart SPM (AIST) are used to determine the perovskite flake thickness.

**Optical simulation**

The optical constants for all 2D HOIP perovskites obtained from the ellipsometry technique reported in our prior publication was taken for the simulation (13). Normal incidence of light on a perovskite flake (similar to the experimental set-up) placed over the Au substrate was used as a simulation model. Using a transfer-matrix method (TMM), the reflected light normal to the substrate was used to compute the reflectance *versus* wavelength for various HOIP thicknesses (36,37). Reflectance from the perovskite flakes deposited on the Au and quartz substrate was compared. It should be noted that a variation (~20%) between the calculated reflectance spectrum for HOIP flake thickness measured experimentally emerges owing to the standard deviation from both the simulation and experiment.

**Acknowledgments**

D.J. acknowledges primary support for this work by the U.S. Army Research Office under contract number W911NF-19-1-0109 and Asian Office of Aerospace Research and Development of the Air Force Office of Scientific Research (AFOSR) FA2386-20-1-4074. S.B.A. gratefully acknowledges the funding received from Swiss National Science Foundation (SNSF) under the Early Postdoc Mobility grant (187977) for this work. H.Z. was supported by Vagelos Institute of Energy Science and Technology graduate fellowship at the University of Pennsylvania. The experiments were carried out at the Singh Center for Nanotechnology at the University of Pennsylvania which is supported by the National Science Foundation (NSF) National Nanotechnology Coordinated Infrastructure Program grant NNCI-1542153. J-C.B and A.D.M acknowledges research support from the HydroGEN Advanced Water Splitting Materials Consortium, established as part of the Energy Materials Network under the U.S. Department of Energy, Office of Energy Efficiency and Renewable Energy, Hydrogen and Fuel Cell Technologies Office, under Award Number DEEE0008843. J.R.H acknowledges support from the Air Force Office of Scientific Research (Program Manager Dr. Gernot Pomrenke) under award number FA9550-20RYCOR059.

## Author contributions

D.J. conceived the idea and supervised the project. S.B.A. implemented the project and performed all steady state optical (PL, reflectance, temperature dependence) characterizations, atomic force microscopy, and data analysis. Time-resolved PL and power dependence PL measurements were recorded by C.E.S under the supervision of J.R.H. The results were analyzed by S.B.A and C.E.S. in discussion with J.R.H. J.L setup the TMM calculations using Python to compute the optical dispersions. B.S. measured optical constants for all HOIP crystals prepared by J.H under the supervision of J-C.B and A.D.M. H.Z and K.J helped in gold evaporation and template-stripped gold substrates, respectively. P.K and H.Z. assisted in setting up the Linkam stage and cryogenic measurements. S.B.A. and D.J. wrote the manuscript with inputs from all co-authors. The authors thank Francisco Barrera for the preliminary work on optical simulation of 2D HOIPs.

**Supporting Information**

**Section 1: Correlating Rabi splitting ($\hbar\Omega$) and extinction coefficient ($k$)**

For strong coupling, the Rabi splitting is a function of coupling strength ($g$), which is related to oscillator strength ($f$) and mode volume ($V_m$) as (38),

$$g = \sqrt{\frac{\pi e^2 f}{4\pi \epsilon m V_m}} \tag{S1}$$

where, $m$ is the free electron mass and $\epsilon$ is the complex frequency-dependent permittivity of the HOIPs. The refractive index ($\epsilon(E)$) of the HOIP can be represented as the sum of a series of Lorentz oscillators as,

$$\epsilon(E) = \epsilon_\infty + \sum_j \frac{f_j}{E_j^2 - E^2 - iE\gamma_j} \tag{S2}$$

where, $\epsilon_\infty$ is the real permittivity at infinite wavelength, $f_j$ is the oscillator strength of the j$^{th}$ oscillator, $E_j$ is the exciton energy, and $\gamma_j$ is the damping factor. However, the permittivity is dominated by the exciton as we approach close to the exciton energy ($E_x$). Therefore, the permittivity as

$$\epsilon(E_x) \approx \frac{f}{E_x^2 - E_x^2 - iE_x\gamma_x} = -\frac{f}{iE_x\gamma_x} \tag{S3}$$

The extinction coefficient ($k(E)$) can then be written as

$$k(E) = \sqrt{\frac{|\epsilon(E)| - Re(\epsilon(E))}{2}} \tag{S4}$$

From equation (S3) and (S4),

$$k = \sqrt{\frac{f}{2E_x \gamma_x}} \qquad (S5)$$

From equation (S1) and (S5),

$$g = k \sqrt{\frac{E_x \gamma_x}{2\epsilon m V_m}} \qquad (S6)$$

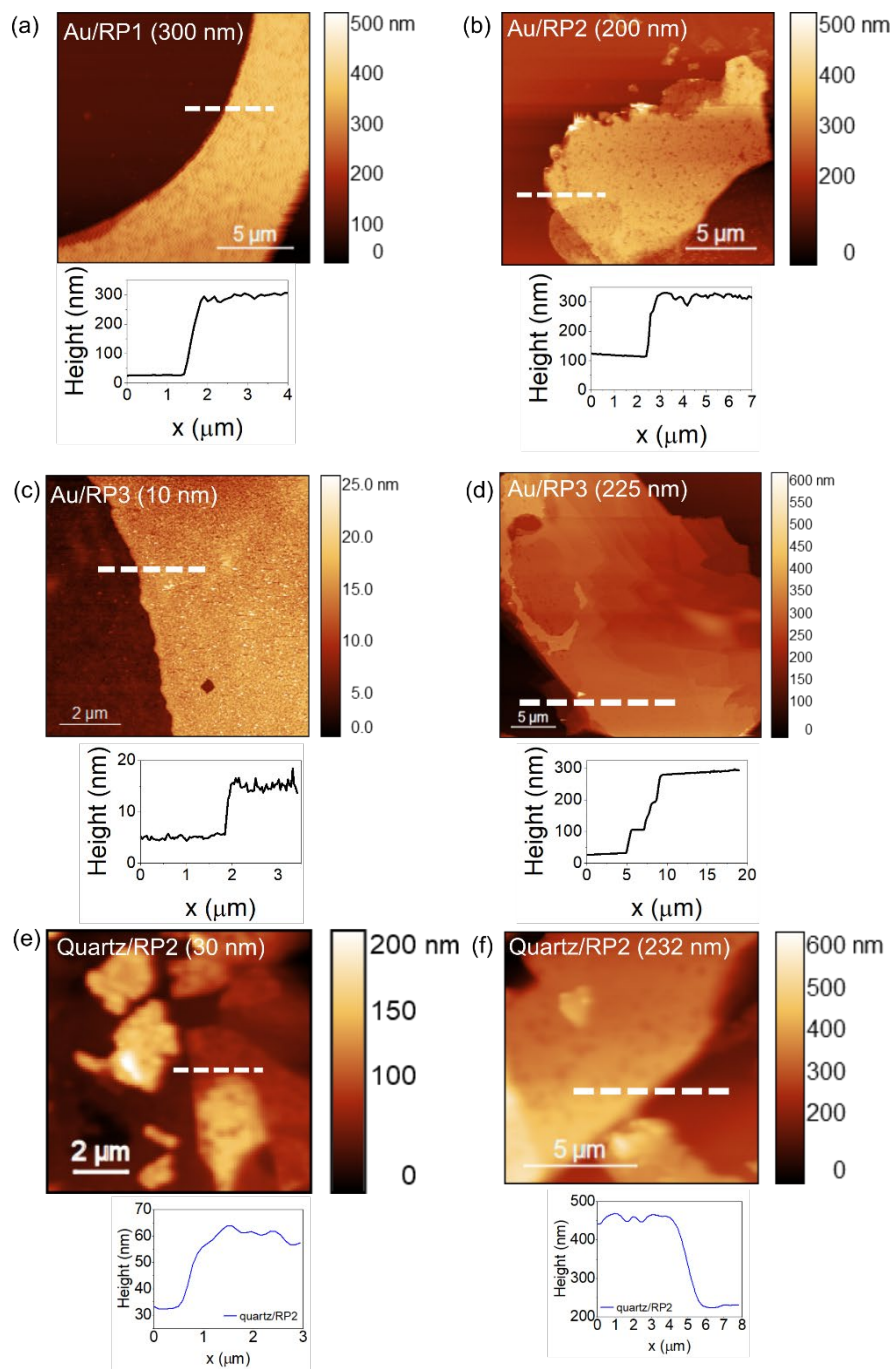

**Figure S1. The topography of the 2D HOIP flakes.** (a-d) Topography of RP1 to RP3 flakes with different thicknesses on Au substrate. (e,f) For comparison, RP2 flakes on the quartz substrate. The line profile (white dotted line) from the topography is shown below their corresponding images.

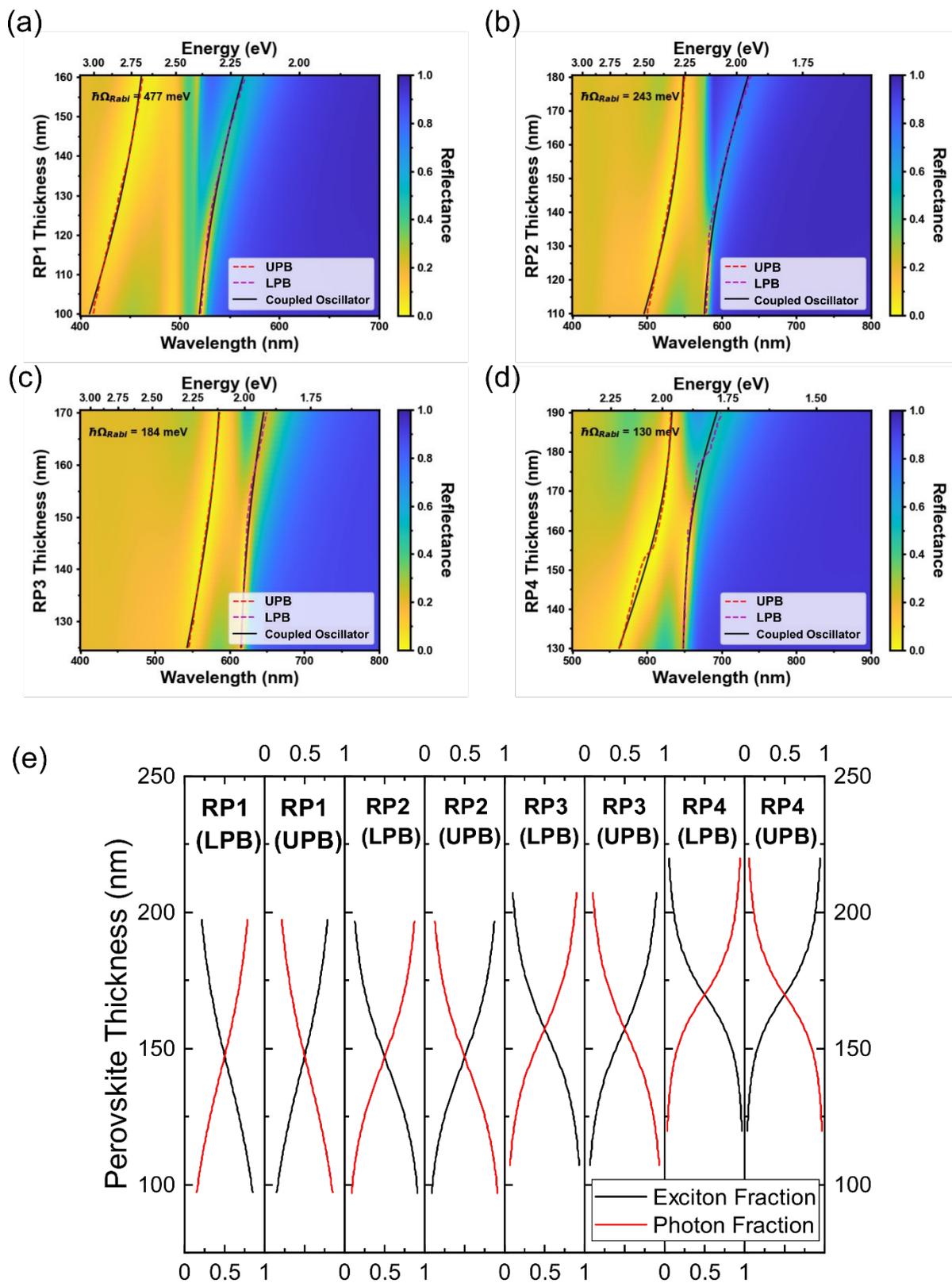

**Figure S2. Two coupled oscillator model fits for HOIP flakes on the Au substrate and Hopfield coefficients.** The reflectance vs. wavelength for different thicknesses of HOIP flakes was fitted with a two coupled oscillator model for (a)

RP1, (b) RP2, (c) RP3, and (d) RP4. The fitted lines shown in solid black lines are used to obtain the upper polariton branch (UPB) and lower polariton branch (LPB). The Rabi splitting calculated from the fitting is mentioned in the inset. (e) Hopfield coefficients (α - exciton fraction, β - photon fraction) for LPB and UPB from all HOIPs. The scale for the coefficients for LPB (UPB) corresponds to the bottom (top) axis.

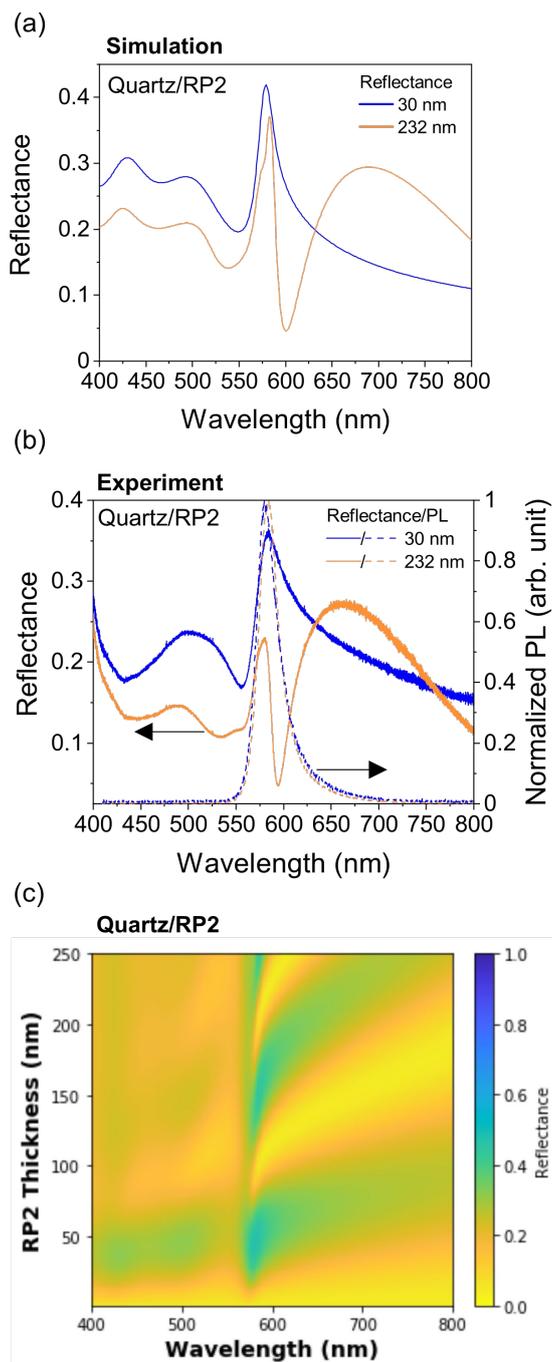

**Figure S3. Optical properties of RP2 flakes on the quartz substrate.** Simulation (a) and experimental (b) reflectance of RP2 (30 nm and 232 nm) on quartz substrate. No significant peak splitting in reflectance and PL was observed. (c) 2D false-color map of calculated reflectance spectra for varying RP2 thickness up to 250 nm on a semi-infinite quartz substrate.

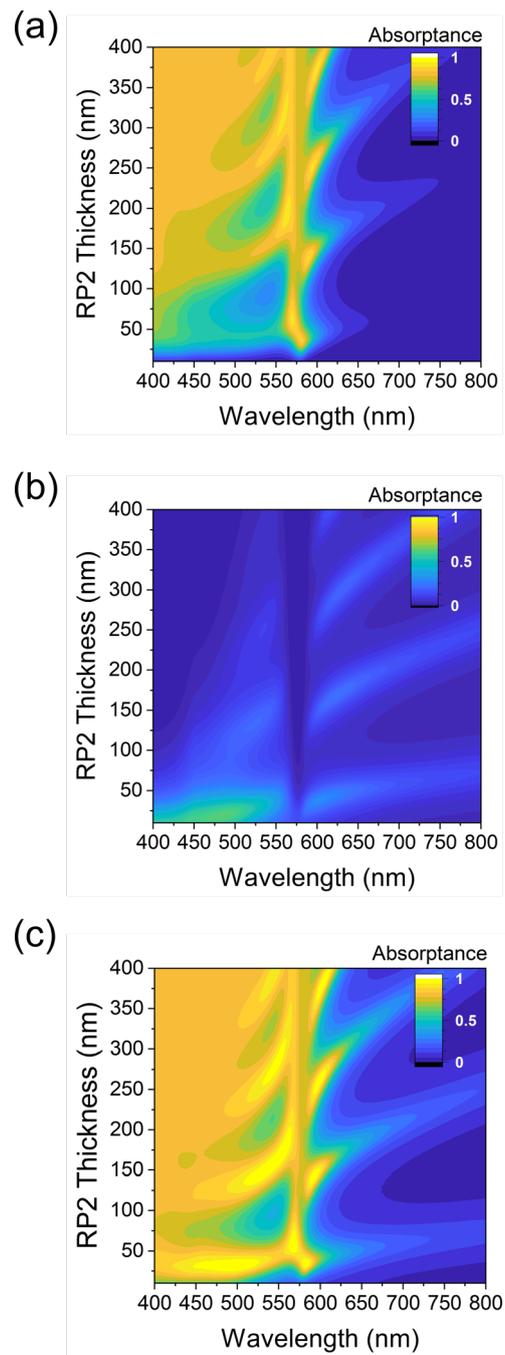

**Figure S4. False color absorption for individual layers and total absorption for RP2 stacked on Au.** Absorption of RP2 (a) and gold (b) for varying RP2 thickness on Au. The total absorption (by adding (a) and (b)) is shown in (c).

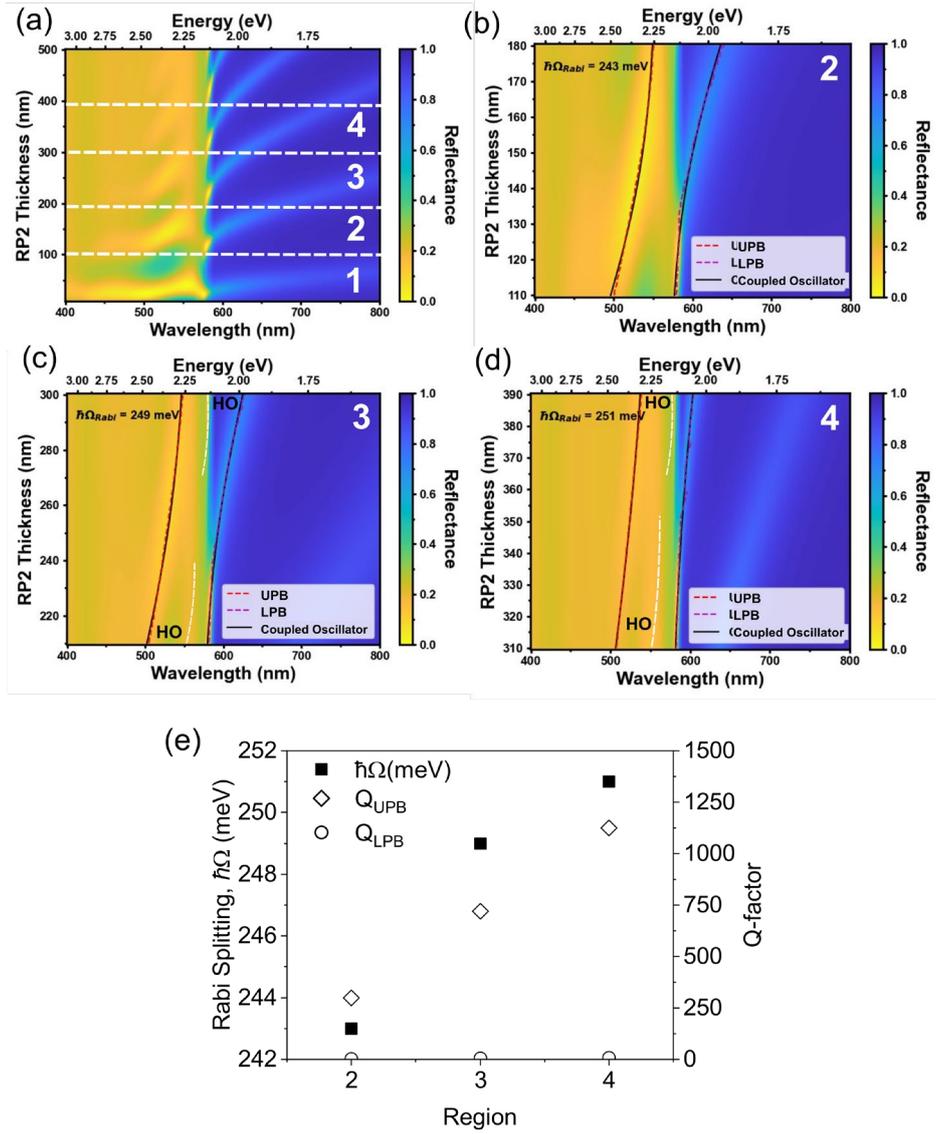

**Figure S5. Rabi splitting from Au/RP2 with different RP2 thicknesses.** (a) The false-color reflectance for Au/RP2 with increasing RP2 thickness up to 400 nm is shown, and the regions are marked from 1-4. Two coupled oscillator model fitting shows clear Rabi splitting in (b) region 2 (110-180 nm), (c) region 3 (210-300 nm), and (d) region 4 (310-390 nm). A slight increase in Rabi splitting with an increase in RP2 flake thickness is ascribed to an increase in the cavity mode resonance. The HO mode is shown in white dotted lines with increasing RP2 thickness in (c) and (d). (e) Rabi splitting and the Q-factor from UPB and LPB resonance were calculated at different regions.

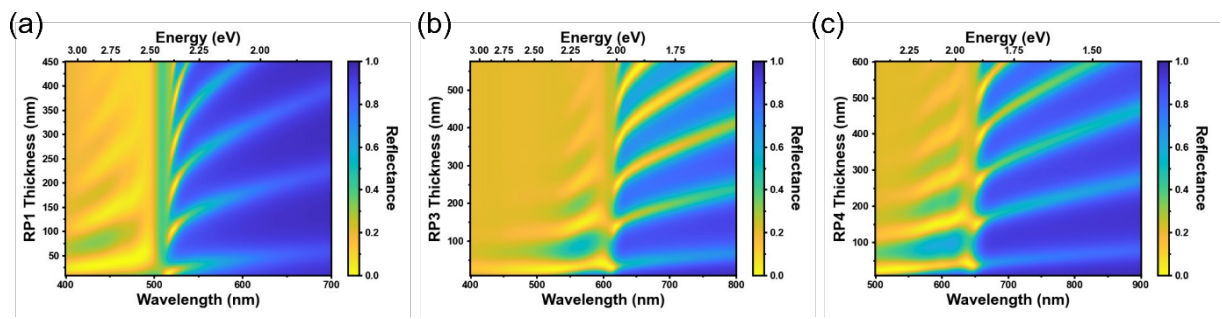

**Figure S6. Thickness-dependent Rabi splitting for RP1, RP3, and RP4 on the Au substrate.** (a-c) False-color reflectance plot showing the anticrossing behavior and emergence of multi-mode lower polariton branches with increasing the thickness of HOIPs (detuning the cavity).

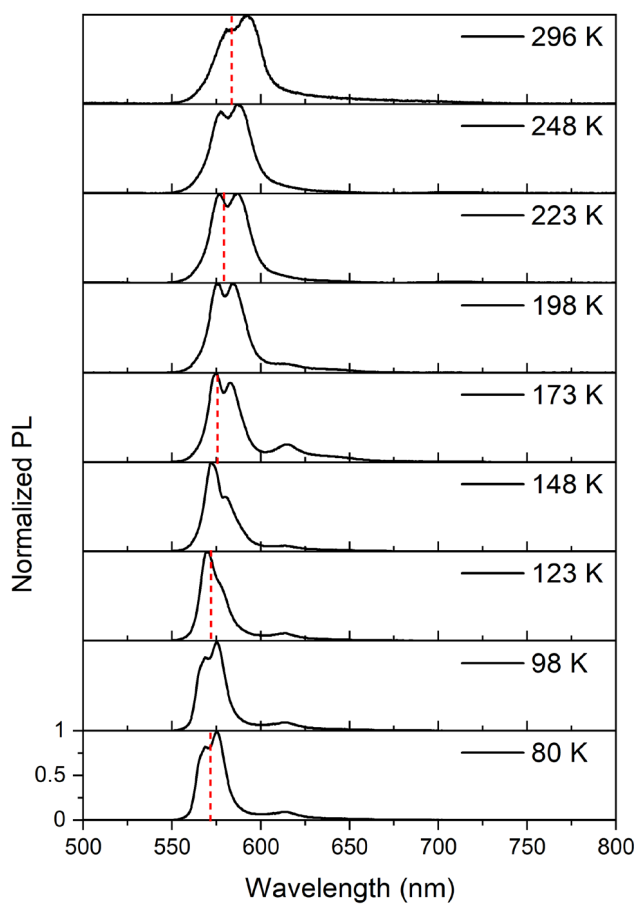

**Figure S7. Temperature dependent normalized PL for 180 nm thick RP2 on the Au substrate.** The normalized PL from 180 nm thick flakes shows hybrid exciton-

polariton emission is persistent even at low temperature (80 K). The normalized PL from 25 nm thin flakes recorded at an interval of 50 K are shown as dotted lines in red.

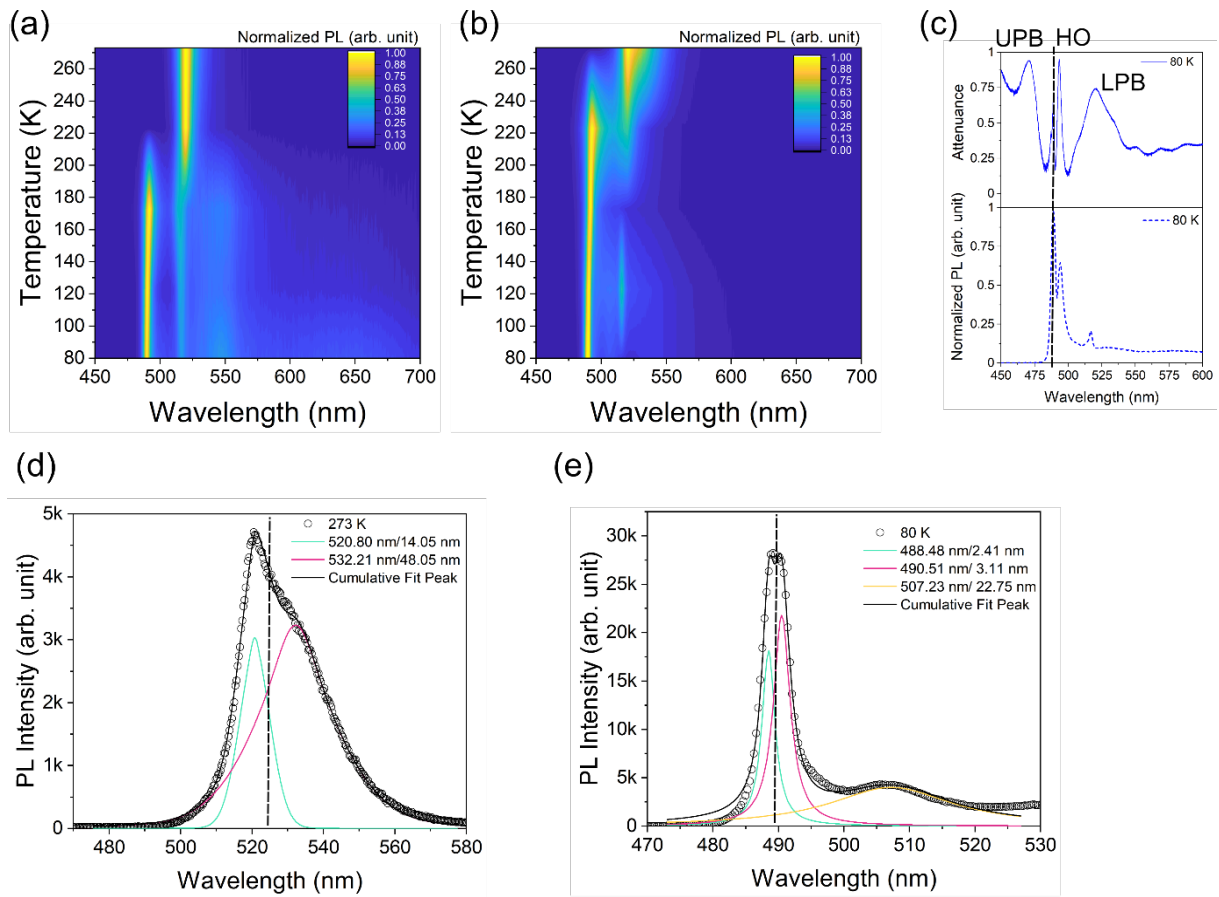

**Figure S8. Temperature-dependent PL from Au/RP1 flakes.** Normalized PL shows a blueshift in the PL peak position upon cooling to 80 K, signifying a phase transition for thin ~15 nm (a) and thick ~ 181 nm flakes (b). (c) Selected attenuance and PL for 181 nm thick flake at 80 K show a clear PL splitting from the HO mode and LPB at lower energy (515 nm). UPB and HO peaks can be observed from PL deconvolution at 296 K (d). In contrast, LPB emission was observed at 80 K (e). The dotted line in (c-e) represents the exciton peak from thin flakes.

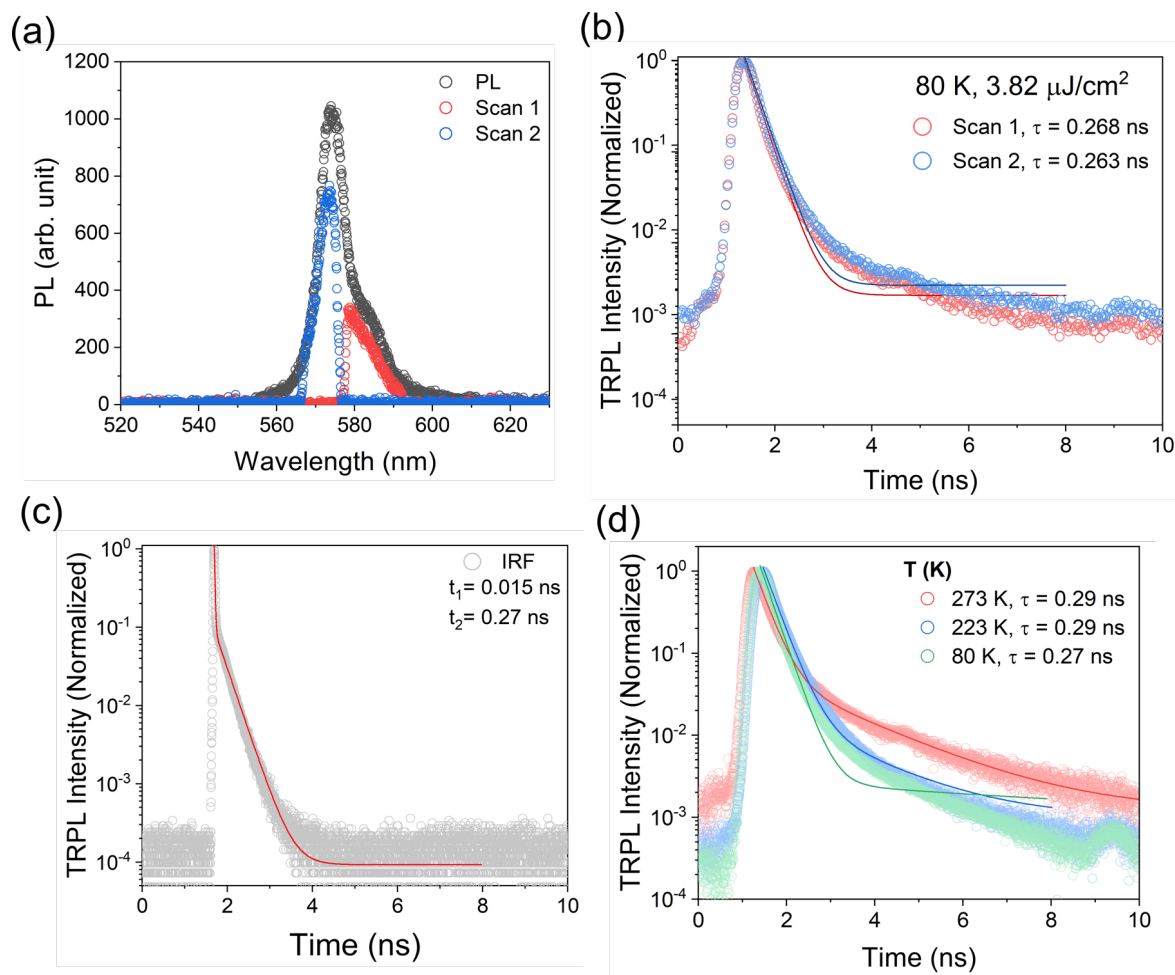

**Figure S9. Time-resolved PL from Au/RP2 (thin flakes) at 80 K.** (a) PL from thin flakes upon excitation with 405 nm and laser power = 3.82 µJ/cm², showing two filtered scans (1 and 2) which are used to obtain TRPL data (b). (c) The instrument response function (IRF) is shown to be faster than the exciton lifetime, as shown in (b). (d) Temperature-dependent exciton lifetime remains unchanged while fitting changes from biexponential at 273 K to monoexponential at 80 K.

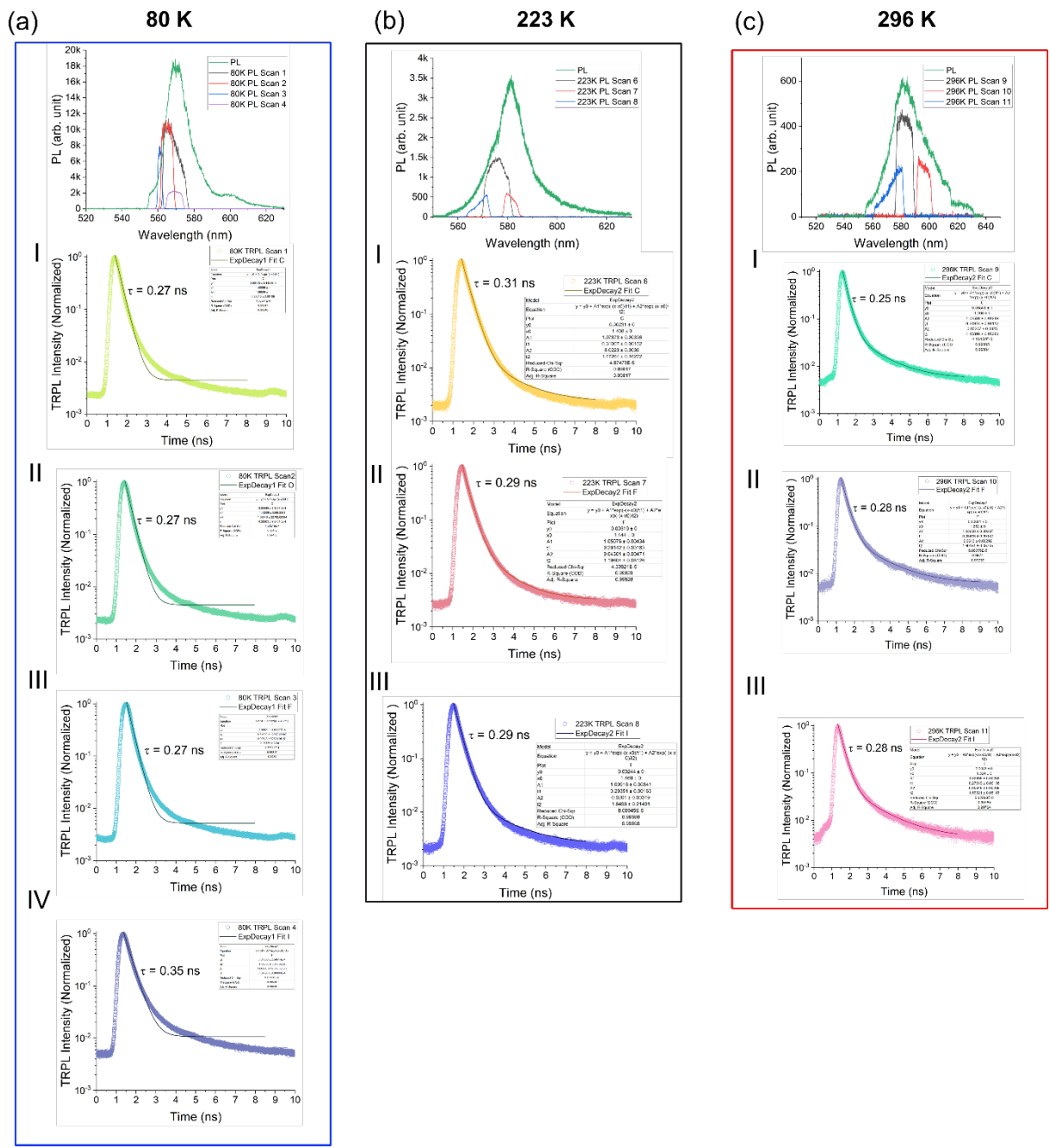

**Figure S10. Time-resolved PL from Au/RP2 (thin flakes) at different temperatures.** PL recorded from thin flakes at 80 K (a), 223 K (b) and 296 K (c). The TRPL from different filtered regions in the PL are fitted with monoexponential fit for 80 K and biexponential fit for 223 K and 296 K as shown in the panels below (I-IV).

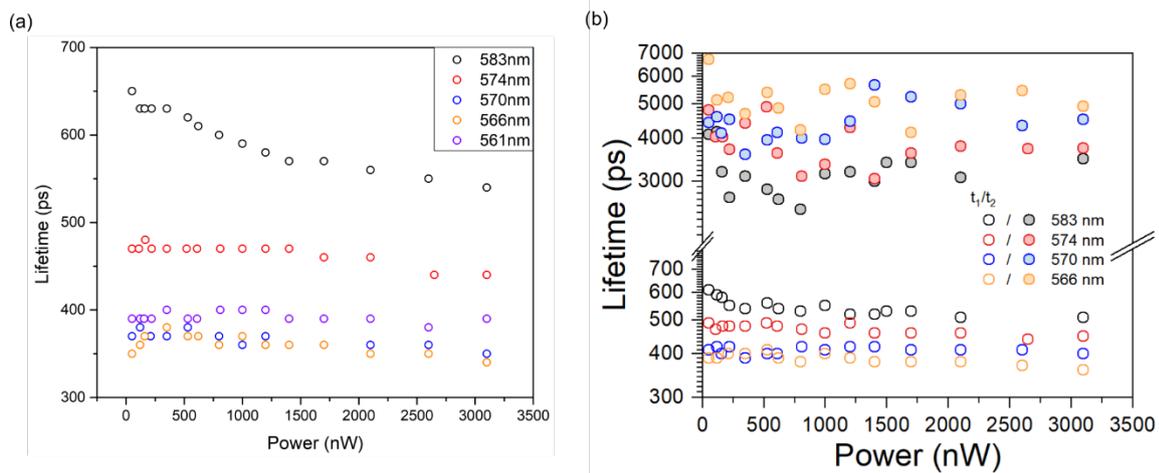

**Figure S11. Exciton lifetime calculated from mono- and bi-exponential fits.** TRPL fitted with a monoexponential (a) and bi-exponential (b) function to calculate the exciton lifetime at different PL filtered wavelength.

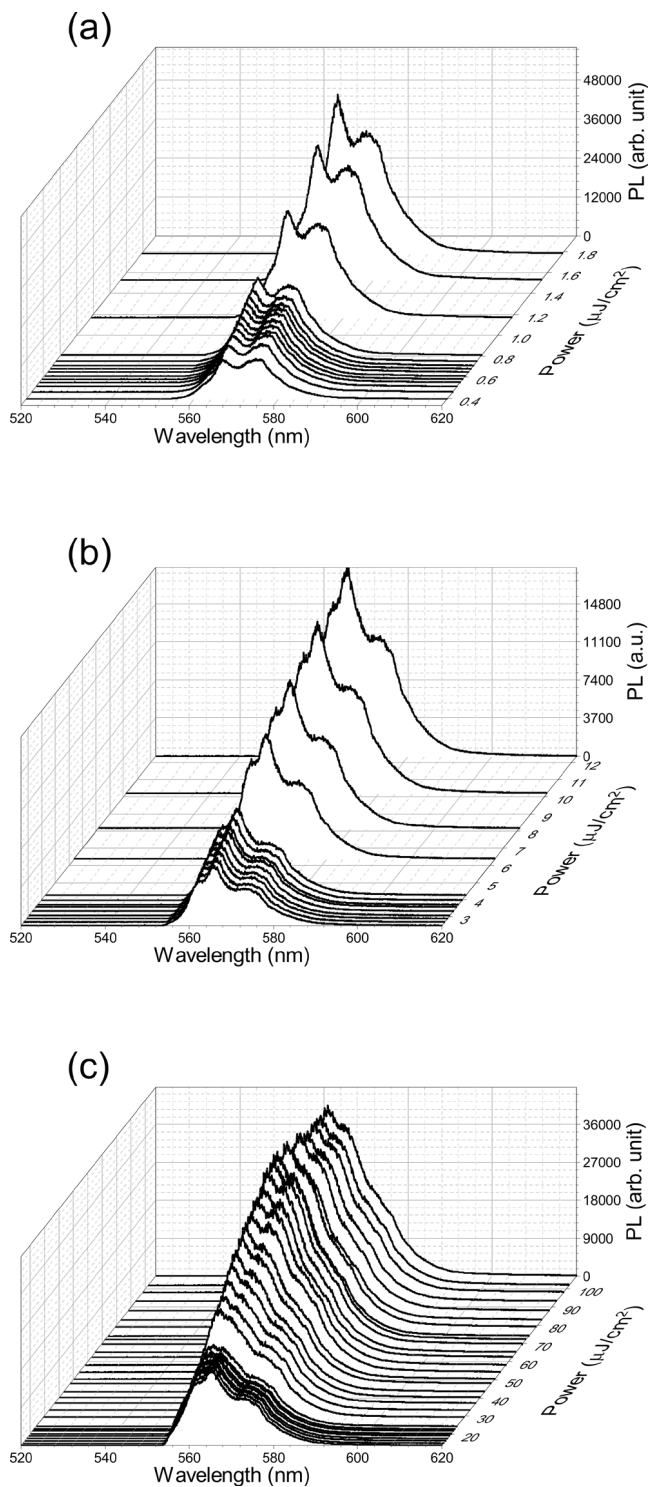

**Figure S12. Power dependent PL from exciton-polariton in Au/RP2 at 80 K.** PL spectra shown for excitation at 405 nm with different power range (a) 0.2- 2 µJ/cm$^2$ , (b) 2-12 µJ/cm$^2$ and (c) 10-110 µJ/cm$^2$.